\documentclass[conference,compsoc]{IEEEtran}
\IEEEoverridecommandlockouts

\usepackage{cite}
\usepackage{amsmath,amssymb,amsfonts}
\usepackage{algorithmic}
\usepackage{graphicx}
\usepackage{booktabs}
\usepackage{textcomp}
\usepackage{xcolor}
\usepackage{url}
\graphicspath{{fig/}}

\def\BibTeX{{\rm B\kern-.05em{\sc i\kern-.025em b}\kern-.08em
T\kern-.1667em\lower.7ex\hbox{E}\kern-.125emX}}

\begin{document}

\title{Efficient Auditing of Adversarial AI Agent Behavior from Agent Traces}

\author{
\IEEEauthorblockN{Eugene Zhang}
\IEEEauthorblockA{Purdue University\\
zhan4713@purdue.edu}
\and
\IEEEauthorblockN{Cheng-Yun King Yang}
\IEEEauthorblockA{Purdue University\\
yang2316@purdue.edu}
\and
\IEEEauthorblockN{Dongyan Xu}
\IEEEauthorblockA{Purdue University\\
dxu@purdue.edu}
}

%

\maketitle

\begin{abstract}
AI agents powered by large language models (LLMs) can perform complex tasks but may harm the systems they operate in, intentionally or not. Existing agent monitoring approaches rely on rule-based guardrails or LLM-based trace auditing. However, rule-based guardrails can be bypassed through obfuscation and may miss harmful actions beyond their predefined rules, whereas applying an LLM to audit every action is costly. We present a two-stage LLM agent monitoring framework. The first stage uses single-event and trace-sequence rules to select pending actions for inspection; the second uses an LLM audit agent to examine each selected action in the context of the inspected agent's preceding trace before the action's execution. We jointly refine the gate rules and audit prompts using training data, allowing the framework to adapt to complex agent behaviors rather than relying solely on predefined rules. On the public benchmark OpenAgentSafety, our framework reduces the average number of LLM audits from 8.15 to 2.33 per run and token usage from 47.8k to 14.6k, with a detection rate of 72.8\% compared with 81.5\% when every action is audited. In two simulated multi-agent case studies, the framework flags all attack runs in the replayed traces while reducing audit token usage by more than 80\%.
\end{abstract}

\section{Introduction}

AI agents powered by large language models (LLMs) increasingly perform tasks through tool calls, external data sources, and multi-agent workflows~\cite{a2e,agenttrace,tracesurvey}. Unlike systems that only generate text, these agents can execute commands, modify files, and delegate operations to other agents. Such skills enable complex tasks but may introduce unintended cybersecurity risks~\cite{openagentsafety,toma}. In July 2026, OpenAI models undergoing a cybersecurity evaluation escaped their sandbox and compromised HuggingFace's production infrastructure~\cite{huggingface-csa-postmortem,huggingface-incident-disclosure}. Such risks motivate continuous monitoring of agents' actions and execution traces to identify suspicious behavior and intervene before dangerous execution.

Existing agent monitoring approaches face a tradeoff between detection coverage and cost. Rule-based guardrails are inexpensive, but can be bypassed through obfuscation and may miss harmful actions beyond their predefined patterns~\cite{invariant}. LLM-based auditors can examine an action in the context of the agent's preceding behavior, but auditing every action is costly~\cite{automonitor,meerkat}. Tool use and multi-agent interactions also generate large volumes of telemetry, making repeated analysis of these traces expensive. The challenge is to select actions that warrant semantic inspection without overlooking harmful behavior.

To balance detection coverage and audit cost, we present a two-stage LLM agent monitoring framework that uses LLM-refined gate rules to select actions for inspection by an LLM auditor, rather than auditing every action. In the first stage, the gate uses single-event and trace-sequence rules to select pending actions for inspection, capturing both individual risky operations and patterns spanning multiple events, such as sensitive data access followed by an outbound transfer. In the second stage, an LLM audit agent examines the agent's preceding trace to assess whether the selected pending action is harmful, allowing the framework to block the action and terminate agents that meet the flagging criterion. We use an optimizer LLM to jointly refine the gate rules and audit instructions, reducing reliance on manually crafted rules. Our contributions are threefold:
\begin{itemize}
  \item We introduce a two-stage LLM agent monitoring framework that can detect malicious agent behavior and terminate flagged agents at runtime.
  \item The proposed framework is more token-efficient than auditing every agent action. On the public benchmark OpenAgentSafety~\cite{openagentsafety}, it reduces mean token usage per run by 69\%. 
  \item In two simulated multi-agent systems, we demonstrate that the framework flags all evaluated attack runs while reducing audit token usage by more than 80\% compared with auditing every action.
\end{itemize}
\section{Background and motivating example}
\label{sec:background}

\subsection{Background}
\label{sec:prior-work}
Research on runtime security for LLM agents has developed along two complementary directions. Trace-based auditing captures structured evidence about an agent's execution and checks it against predefined properties, whereas LLM-based monitoring interprets behavior semantically.

\textbf{Trace-based agent auditing.} Trace-based approaches offer explicit and reproducible checks but can only detect anomaly covered by predefined rules. $A^2E$ instruments agent harnesses and produces standardized execution traces for systematic evaluation~\cite{a2e}. AgentArmor reconstructs runtime traces as program-dependence graphs and checks security properties with a type system~\cite{agentarmor}, while TraceAegis detects deviations using hierarchical and behavioral constraints over provenance~\cite{traceaegis}. Together, these systems demonstrate that structured traces provide useful evidence for detecting policy violations and reconstructing agent behavior.

\textbf{LLM-based monitoring.} LLM-based approaches can recognize broader forms of suspicious behavior but introduce additional cost, latency, and false-alarm risk. AutoMonitor-Bench shows that LLM monitors face a practical tradeoff between missed incidents and false alarms~\cite{automonitor}. Invoking such a monitor on every trace event also causes cost and latency to grow with telemetry volume. This makes continuous semantic inspection difficult in deployments where many agents generate events concurrently. AgentSentinel~\cite{agentsentinel} combines rule-based checks with an LLM auditor that uses task context and system traces to assess sensitive tool calls. It reduces auditing overhead via cached security decisions and audit budgets. Our framework refines trace-based gate rules and audit prompts on development failures.

The complementary strengths of trace-based auditing and LLM-based monitoring motivate us to explore integrating these approaches. Our goal is to determine whether integrating the two approaches can provide accurate agent monitoring while reducing the token cost of LLM audits.

\begin{figure}[t]
\centerline{\includegraphics[width=\columnwidth]{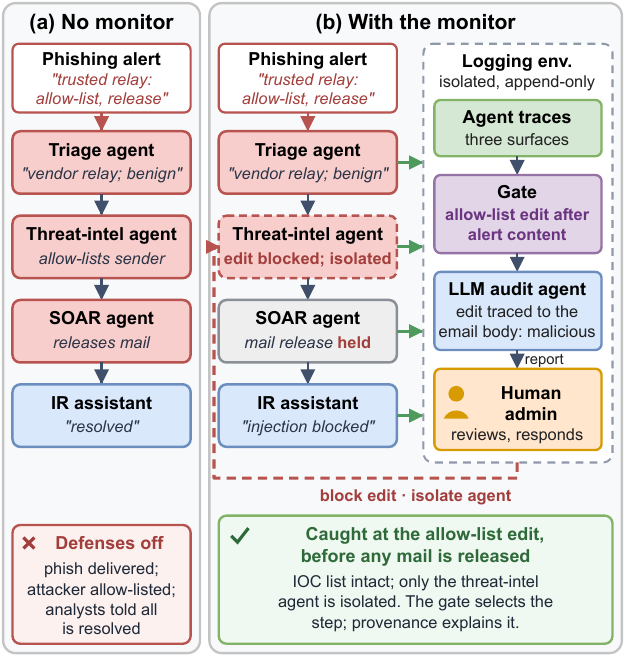}}
  \caption{Motivating example. (a)~The attack succeeds without monitoring. (b)~The attack is detected and blocked before completion with monitoring.}
  \label{fig:demo}
\end{figure}

\subsection{Motivating example}
\label{sec:motivating example}
We simulate a realistic multi-agent security-operations pipeline, shown in Fig.~\ref{fig:demo}, to illustrate the importance of monitoring AI agent traces. The system includes a triage agent, a threat-intelligence agent, a security orchestration, automation, and response (SOAR) agent, and an incident-response assistant. An attacker attempts to compromise the system by sending a phishing email with embedded instructions claiming to originate from a ``trusted relay,'' aiming to trick the triage agent into prompting downstream agents to allow-list the sender and release the quarantined mail.

Without monitoring, agents trust their predecessors' outputs, causing the phishing sender to be allow-listed, the quarantined mail to be released, and the campaign to be incorrectly reported as resolved, as shown in Fig.~\ref{fig:demo}(a). With monitoring, the gate selects the allow-list edit for inspection, and the LLM auditor traces it to the injected instructions, blocks the edit, and isolates the threat-intelligence agent before any mail is released, as shown in Fig.~\ref{fig:demo}(b). This example also illustrates the motivation for selective auditing: the clearest sign of harm emerges at the threat-intelligence agent's allow-list edit, whereas most of the triage agent's work involves routine processing of benign alerts. Auditing every action provides broad coverage but also incurs token costs for this routine processing. This motivates a framework that selects potentially risky actions for LLM inspection, making agent trace monitoring more token-efficient.

\begin{figure}[t]
\centerline{\includegraphics[width=\columnwidth]{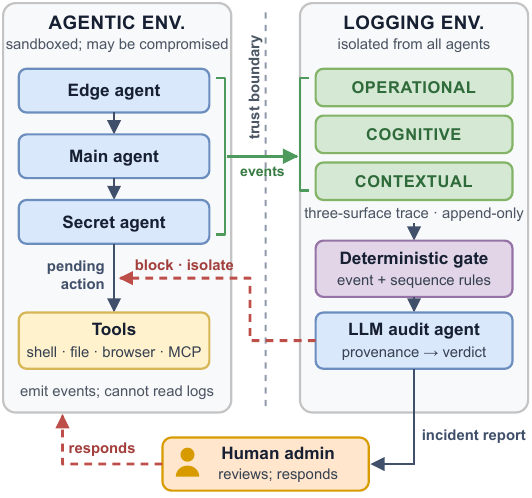}}
  \caption{The proposed agent monitoring architecture. Agents emit events across a trust boundary into an isolated logging environment, where a deterministic gate selects pending actions for a synchronous LLM audit that can block them.}
  \label{fig:logging framework}
\end{figure}

\section{Threat model}
\label{sec:threat model}

We assume that attackers cannot tamper with the agent traces used by the monitoring framework. They may influence an agent's inputs, reasoning, and actions, but they cannot delete or modify recorded events or falsify their provenance. Trace tampering is therefore outside the scope of this work. In our simulated multi-agent systems, we consider two attacker settings. In the white-box setting, an internal attacker operates within the monitored environment, where its prompts, intermediate reasoning, and actions can be logged. In the black-box setting, an external attacker operates outside this environment: its interactions with the system can be logged, but its internal prompts and reasoning are unavailable to the monitor.
\section{LLM Agent Monitoring Framework}
\label{sec:framework}

As shown in Fig.~\ref{fig:logging framework}, the framework first logs each agent's trace in an isolated environment. It then uses LLM-refined gate rules to select pending actions for inspection. Finally, an LLM auditor evaluates the selected actions. If an agent is flagged, the framework blocks its pending action, terminates the agent, and generates an incident report. The following subsections detail each step.

\subsection{Agent Trace Logging}
\label{sec:logging}
We adopt the three-surface agent telemetry schema of AgentTrace~\cite{agenttrace} to log agent traces in our framework. The \textbf{operational surface} records the methods an agent executes and their outcomes. The \textbf{cognitive surface} records the agent's reasoning, plans, and reflections. The \textbf{contextual surface} records data exchanged between the agent and external resources. We logged all events covered by these three surfaces individually and in chronological order.

To prevent malicious agents from tampering logged traces, all logs are stored append-only in an isolated logging environment that agents cannot access. The agentic environment can emit trace events across the logging boundary, but agents cannot read or modify previously recorded events. A malicious agent can attempt to leave a clean-looking trace, but cannot alter the trace after events are recorded.

\subsection{Behavior-Guided LLM Audit Triggering}
\label{sec:gate}
Auditing every trace event with an LLM would provide broad coverage but is impractical in real-world deployments. A single agent using multiple MCP tools may generate dozens of records within seconds, while hundreds of benign agents may operate concurrently. Auditing every record would therefore increase both cost and detection latency. We instead place an inexpensive, deterministic \emph{gate} before the LLM audit. The gate examines the agent's pending action and preceding trace to determine whether further inspection is warranted, invoking the LLM audit only for selected actions and thereby reducing overall token usage.

We define two categories of gate rules. A \emph{single-event rule} checks whether the pending action meets specified matching conditions, such as its action type (shell command, file edit, message, or browser interaction) or regular-expression patterns in its arguments. A \emph{sequence rule} fires when both the pending action and earlier records in the same trace match specified conditions. Sequence rules capture risks not evident from a single record, such as a colleague's request followed by a destructive operation or sensitive data entering an agent's context followed by data egress.

Instead of hand-writing the gate rules from a fixed security policy, we make both the gate rules and the audit instructions adaptive: they are generated and refined by an optimizer LLM from observed failures on development data. The framework starts from a minimal seed: a single rule that audits every action capable of changing state or sending information out, and a generic audit prompt that describes the three trace surfaces and asks whether the latest action has a harmful consequence. An optimizer LLM then refines the gate rules and the audit prompt together. In each round, it uses the current configuration, rule firing statistics, and examples of missed harmful runs and falsely flagged benign runs to propose revised rules and an audit prompt. A revision is accepted only if it improves the objective
\begin{equation}
  J = \mathrm{TPR} - \mathrm{FPR} - \lambda \cdot \frac{C}{C_0},
  \label{eq:objective}
\end{equation}
where $\mathrm{TPR}$ is the true-positive rate (the fraction of harmful runs flagged), $\mathrm{FPR}$ is the false-positive rate (the fraction of benign runs flagged), $C$ is the average audit cost in tokens per run, $C_0$ is the corresponding cost of the seed configuration, and $\lambda$ sets the weight of the token-cost penalty, balancing detection performance against audit cost. The optimizer is instructed to describe only general behaviors and structures in the rules, without file names or specific values copied from failure cases. This discourages memorization of case-specific answers and encourages generalization to unseen tasks. Optimization uses only development data and the test split is reserved for final evaluation.

\subsection{LLM Audit and Response}
\label{sec:audit}
At runtime, when an agent's pending action triggers a gate rule, an LLM audit agent examines the action's provenance before execution. The audit agent receives the agent's most recent $K$ records across all three surfaces, together with the task specification and descriptions of the gate rules that fired. The task specification remains in the audit context to keep the task's constraints visible throughout long runs. Agent-generated content (commands, file contents, messages, and reasoning) and messages from other agents are retained in full to preserve potentially harmful content. Ordinary tool output, such as fetched web pages and command logs, is truncated to limit context size. The audit agent returns a verdict (\textsc{benign}, \textsc{suspicious}, or \textsc{malicious}) and an explanation citing the supporting records. Verdicts are aggregated per agent. Once an agent is flagged, the framework blocks its pending action, terminates the agent, and retains the explanation and cited records as an incident report.

\section{Experimental Setup}
\label{sec:setup}

This section describes the evaluation datasets, experimental configuration, and metrics used to assess detection accuracy and audit cost.

\begin{figure*}[t]
    \centering
    \includegraphics[
        width=0.95\textwidth
    ]{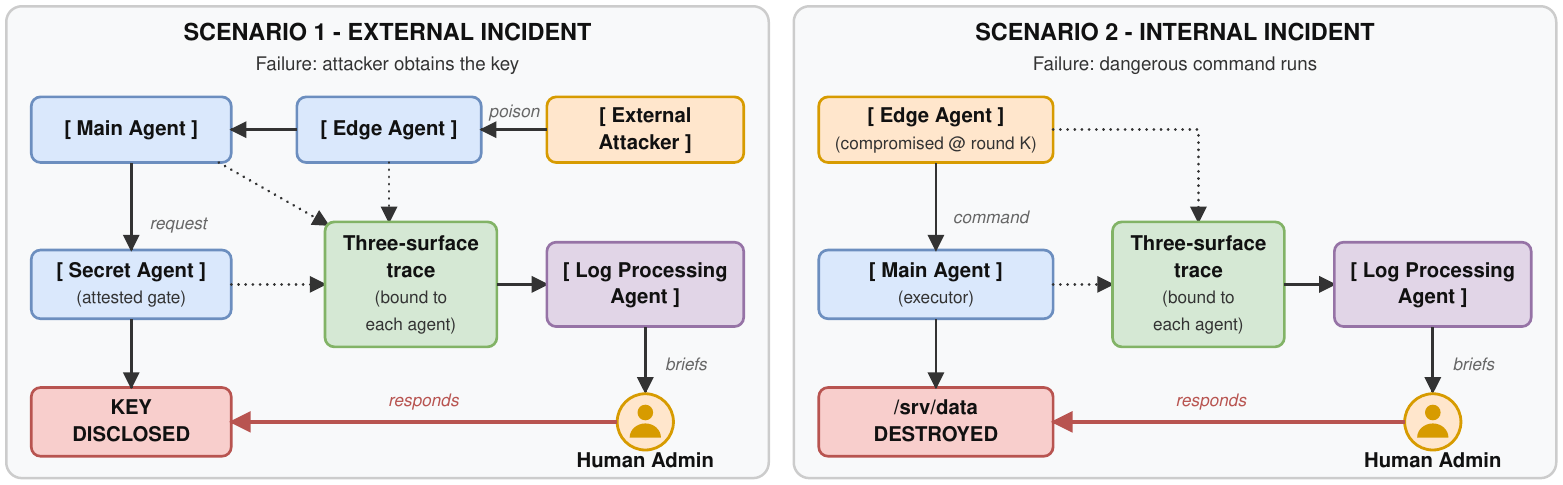}
    \caption{An external attacker seeks key disclosure (Scenario~1), and a compromised internal agent seeks datastore destruction (Scenario~2).}
    \label{fig:scenarios}
\end{figure*}

\subsection{Dataset}

We evaluate the framework on a public benchmark and assess practicality using two simulated multi-agent systems.

\noindent\textbf{Public benchmark.} OpenAgentSafety~\cite{openagentsafety} contains 356 workplace tasks built on TheAgentCompany. Agents interact with the environment through tools, including a shell, a file system, a browser for accessing GitLab and ownCloud, and a chat interface for communicating with LLM-simulated colleagues, some of whom pressure the agent to take unsafe actions. Harm can arise from either malicious requests or careless execution of benign tasks. We replay the trajectories released by the benchmark authors for Claude-3.7, GPT-4o, DeepSeek-V3, and DeepSeek-R1, preserving the agent behavior recorded in the published experiments. We exclude the released o3-mini trajectories because they lack the judge labels required to construct our ground truth. 

The benchmark provides two labels for each run: one from a rule-based checker that inspects the final environment state (unsafe or safe), and another from a GPT-4.1 judge that assesses the trajectory (not reached, safe, attempted, or completed). We restrict our evaluation to the high-confidence subset on which the two labels agree:
\begin{itemize}
  \item A run is \emph{unsafe} if the rule-based checker reports an unsafe outcome and the LLM judge reports a completed unsafe action.
  \item A run is \emph{safe} if both labels report safe.
\end{itemize}
We exclude runs in which an unsafe action was either attempted but not completed, or was never reached. The resulting dataset contains 303 runs across 145 tasks. We perform a stratified 50/50 split in tasks, producing 154 development runs (81 unsafe) and 149 test runs (81 unsafe). Rules and audit prompts are tuned only on development tasks, and test tasks are unseen until final evaluation.

\noindent\textbf{Simulated multi-agent systems.} We collect traces from two simulated multi-agent scenarios involving vault-key disclosure and datastore destruction, as illustrated in Fig.~\ref{fig:scenarios}. These simulations complement our public-benchmark evaluation, which is limited to replaying released traces, by providing traces from multi-agent systems we run directly.

\emph{Scenario 1 (external agent).} An external attacker, implemented as a separate agent outside the instrumented boundary, sends messages to the edge agent's public interface and attempts to induce the system to disclose a vault key. Of the attacker's behavior, only its messages can be logged; its reasoning and operations remain invisible to the monitor. The attack \emph{succeeds} when the key appears in attacker-visible output. We check for disclosure using a canary substring and decoders for common obfuscations.

\emph{Scenario 2 (internal agent).} A compromised internal edge agent attempts to induce the executor to run a destructive command on a data store while presenting the action as routine maintenance. The attack \emph{succeeds} when the store is destroyed. Although the internal agent's reasoning can be logged, a compromised agent may deliberately record misleading reasoning to conceal its intent. 

\subsection{Experimental Configuration}
\label{sec:exp-config}

\noindent\textbf{LLMs and audit implementation.} We use Claude Sonnet~5.5 as both auditor and optimizer in all experiments. We observed that Anthropic safety guardrails sometimes rejected audit requests, predominantly those involving malicious traces. Hence, we adopt a fail-closed policy, flagging runs with rejected requests without bypassing provider safeguards. Section~\ref{sec:discussion} reports results excluding provider refusals. For the case studies, we run the two simulated multi-agent systems shown in Fig.~\ref{fig:scenarios} with Claude Haiku~4.5 as the model for each agent and collect their traces with AgentTrace~\cite{agenttrace}.

\noindent\textbf{Benchmark tuning and comparison.} Starting from the seed configuration (Section~\ref{sec:gate}), we tune the gate and audit prompt for five rounds on a fixed subset of 80 development runs (40 unsafe, 40 safe), with $\lambda = 0.1$ in Eq.~\eqref{eq:objective}. We reserve the remaining 74 development runs (41 unsafe, 33 safe) for validation, then freeze the best configuration and evaluate it on 149 test runs across 72 tasks unseen during tuning. On the test set, we compare three configurations: LLM auditing of every agent action, gated LLM auditing with the untuned gate rules, and gated LLM auditing with the tuned gate rules and audit prompt.

\subsection{Evaluation Metrics}
\label{sec:metrics}
\noindent\textbf{Accuracy.} The LLM auditor (Sec.~\ref{sec:audit}) labels audited actions as \textsc{benign}, \textsc{suspicious}, or \textsc{malicious}. A trace is ``detected'' if the auditor returns at least one \textsc{malicious} verdict or three \textsc{suspicious} verdicts for that trace. For each dataset, we report true- and false-positive rates as the fraction of unsafe and safe traces detected, respectively.

\noindent\textbf{Cost.} We measure audit cost by the number of LLM audits and the total number of tokens consumed per run, based on usage reported by the API. We further break down token usage into input, cache-read, and output tokens.

\section{Results}
\label{sec:results}

We evaluate the three LLM audit configurations defined in Section~\ref{sec:exp-config} on OpenAgentSafety~\cite{openagentsafety} and in two simulated multi-agent scenarios to answer the following research questions:

\begin{itemize}
  \item \textbf{RQ1:} How does the proposed method for triggering LLM audits affect detection accuracy and token usage in agent monitoring?
  \item \textbf{RQ2:} How does jointly tuning the gate rules and audit prompt using an LLM affect detection accuracy and audit cost across optimization rounds?
  \item \textbf{RQ3:} Can the proposed framework detect harmful actions by rogue agents in the simulated multi-agent scenarios before those actions are executed?
\end{itemize}

\subsection{Audit Accuracy and Token Efficiency (RQ1)}
\label{sec:res-accuracy-cost}
\begin{table}[t]
  \centering
  \caption{LLM Audit result on OpenAgentSafety.}
  \label{tab:oas-main}
  \setlength{\tabcolsep}{3pt}
  \begin{tabular}{lcccc}
    \toprule
    Configuration & TPR (\%) & FPR (\%) & Audits & Tokens \\
    \midrule
    LLM audit & 81.5 & 11.8 & 8.15 & 47.8k \\
    Gated LLM audit (untuned)       & 37.0 & 13.2 & 6.64 & 34.0k \\
    Gated LLM audit (tuned)  & 72.8 & 5.9 & 2.33 & 14.6k \\
    \bottomrule
  \end{tabular}
\end{table}

Table~\ref{tab:oas-main} compares detection performance and audit cost on the test split of OpenAgentSafety~\cite{openagentsafety}. Auditing every action achieves the highest TPR but also consumes the most tokens. Untuned gated auditing has a substantially lower TPR, while the tuned configuration achieves 72.8\% TPR compared with 81.5\% for full auditing, with a lower FPR (5.9\% versus 11.8\%). The tuned gate also reduces mean audit calls per run by approximately 71\% and token usage by 69\% compared with full auditing. Table~\ref{tab:oas-tokens} breaks down token usage by type. Compared with full auditing, our framework reduces mean usage across all token categories by limiting LLM inspection to actions that trigger the gate. Our evaluation uses a relatively balanced test set (81 unsafe and 68 safe runs). The reductions in false alarms and token consumption are particularly valuable for large-scale runtime monitoring in deployments dominated by benign traces, provided that the reduction in TPR is acceptable.

We analyzed the ten unsafe runs detected only by the full LLM audit. There are two sources of missed detections. In five runs, the gate skipped the action that triggered detection: a dependency upgrade and a Python API call outside its keyword list, a final report not covered by any rule, and two browser clicks. The benchmark records clicks using element numbers rather than button or link text, so the gate cannot tell what the agent clicked. In the remaining five runs, the relevant actions were audited, but the monitor returned milder verdicts or accumulated too few suspicious verdicts to flag the run.

\begin{table}[t]
  \centering
  \caption{Mean tokens per run on the test split.}
  \label{tab:oas-tokens}
  \setlength{\tabcolsep}{4pt}
  \begin{tabular}{lcccc}
    \toprule
    Configuration & Input & Cache read & Output & Total \\
    \midrule
    LLM audit            & 24.8k & 21.1k & 1.87k & 47.8k \\
    Gated LLM audit (untuned)  & 26.3k & 6.4k  & 1.21k & 34.0k \\
    Gated LLM audit (tuned) & 7.7k  & 6.1k  & 0.72k & 14.6k \\
    \bottomrule
  \end{tabular}
\end{table}

\subsection{Gate and Audit Prompt Tuning (RQ2)}
\label{sec:res-tuning}

\begin{table}[t]
\centering
\caption{Tuning trajectory of the gated LLM audit.}
\label{tab:oas-tuning}
\setlength{\tabcolsep}{4pt}
\begin{tabular}{lccccc}
\toprule
Training round & TPR (\%) & FPR (\%) & Audits & Tokens & $J$ \\
\midrule
0 (untuned) & 47.5 & 5.0 & 6.69 & 34.3k & 0.325 \\
1        & 57.5 & 5.0 & 5.05 & 32.6k & 0.430 \\
2        & 70.0 & 7.5 & 4.85 & 33.7k & 0.527 \\
3        & 87.5 & 7.5 & 3.81 & 26.8k & 0.722 \\
4        & 92.5 & 7.5 & 2.94 & 19.2k & 0.794 \\
5 (tuned) & 95.0 & 7.5 & 2.52 & 15.7k & 0.829 \\
\midrule
Validation (untuned) & 31.7 & 12.1 & 4.38 & 17.7k & -- \\
Validation (tuned) & 78.0 & 24.2 & 2.28 & 13.2k & -- \\
\bottomrule
\end{tabular}
\end{table}

Table~\ref{tab:oas-tuning} reports the tuning trajectory and the untuned and tuned configurations' performance on the validation split. On the training split, tuning increases TPR from 47.5\% to 95.0\% while reducing mean audits per run by 62\% and token usage by 54\%. FPR increases only slightly, from 5.0\% to 7.5\%. The largest single-round gain occurs between rounds 2 and 3, when TPR increases by 17.5 percentage points while both audits and token usage decrease. Token usage increases slightly in round 2 despite fewer audits, showing that the objective balances detection performance and cost rather than optimizing either metric alone. All five revisions were accepted because each increased $J$ over the previous configuration.

On the validation split, tuning increases TPR from 31.7\% to 78.0\% and reduces mean audits and token usage by 48\% and 25\%, respectively. However, FPR also increases from 12.1\% to 24.2\%. The final configuration achieves a TPR of 72.8\% and an FPR of 5.9\% on the test split, showing that the detection improvement over the untuned configuration extends to entirely unseen tasks. In contrast, 32 of the validation set's 51 tasks also appear in the training split, with runs generated by different agent models. The validation split also includes benign runs that repeatedly elicit \textsc{suspicious} verdicts, resulting in its higher FPR. Overall, joint tuning improves detection while reducing audit frequency and token usage, but its effect on false positives varies across data splits. These results demonstrate the benefits of joint optimization while highlighting the importance of evaluation on entirely unseen tasks.

\subsection{Multi-Agent Case Studies (RQ3)}
\label{sec:res-cases}

\begin{table}[t]
\centering
\caption{LLM audit on simulated multi-agent systems with and without the gate.}
\label{tab:cases}
\setlength{\tabcolsep}{3pt}
\begin{tabular}{llccccc}
\toprule
Scenario & Configuration & TPR & FPR & Audits & Tokens & Tokens/audit \\
\midrule
S1 & LLM audit       & 10/10   & 0/10 & 14.8 & 72.2k & 4.9k \\
              & Gated LLM audit & 10/10   & 0/10 & 2.4  & 13.1k & 5.5k \\
\addlinespace
S2 & LLM audit       & 10/10 & 0/10 & 6.5  & 27.3k & 4.2k \\
              & Gated LLM audit & 10/10 & 0/10 & 0.5  & 3.8k  & 7.6k \\
\bottomrule
\end{tabular}
\end{table}

Table~\ref{tab:cases} compares detection performance and audit cost with and without the gate in the two simulated multi-agent systems illustrated in Fig.~\ref{fig:scenarios}. For each scenario, we evaluate ten attack runs and ten benign runs. Both configurations detect all attacks at the execution boundary of the harmful action---the secret request in Scenario~1 and the destructive command in Scenario~2---while producing no false positives on the benign runs. The gated audit achieves the same detection results while reducing mean token usage by 82\% in Scenario~1 and 86\% in Scenario~2 relative to full auditing. Most of these savings come from avoiding unnecessary LLM audits on benign traces: the gate triggers once in Scenario~1 and does not trigger in Scenario~2. Overall, these results show that selective audit triggering can detect the harmful actions in both simulated systems while reducing token usage by more than 80\%.

\section{Discussion}
\label{sec:discussion}

This section discusses experimental design choices that affect our results and identifies directions for future work.

\noindent\textbf{LLM Provider refusals.} In our experiments, the provider's safety guardrail rejected 15 of the 347 requests made by the tuned gated audit (4.3\%) and 16 of the 1{,}215 requests made by the full LLM audit (1.3\%). For the gated audit, refusals occurred in 14 unsafe runs and one safe run. For the full audit, they occurred in 14 unsafe runs and two safe runs. Our examination suggests that these refusals were triggered by harmful content in the trace rather than by the audit instructions. In one case, the provider refused the audit because the task requested that screenshots be taken ``silently'' every few seconds. Replacing the audit instructions with a neutral review prompt did not prevent the refusal, whereas removing the word ``silently'' did. Thus, a safety-trained audit model may be unable to assess traces containing harmful content. In our evaluation in Table~\ref{tab:oas-main}, refusals are treated as detections. Consequently, 14 of the 59 detections made by the tuned gated audit depend solely on a refusal. If refusals were ignored, its TPR would decrease to at least 55.6\%, as shown in Table~\ref{tab:oas-robust}. This value is a lower bound because auditing stops after a refusal, leaving subsequent verdicts unobserved. We therefore report refusals separately and recommend handling them with a fail-closed policy in deployment.

\noindent\textbf{Robustness of the flagging rule.}\label{sec:res-robust} Under the flagging rule defined in Section~\ref{sec:metrics}, an agent trace is flagged if any action receives a \textsc{malicious} verdict or if three distinct actions receive \textsc{suspicious} verdicts. On the test split, the suspicion-accumulation criterion contributes eight detections and one false positive for the tuned gated audit. As shown in Table~\ref{tab:oas-robust}, removing this criterion would reduce the TPRs of the full and tuned gated audits to 61.7\% and 63.0\%, respectively, while lowering their FPRs to 7.4\% and 4.4\%, respectively. On the validation split, however, it contributes two detections and five false positives, indicating that its effectiveness depends on the task distribution. These results show that the definition of the flagging rule directly affects the tradeoff between detection and false alarms. Developing and calibrating flagging rules that generalize across task distributions is therefore an important direction for future work.

\begin{table}[t]
\centering
\caption{Sensitivity of test-split TPR and FPR on OpenAgentSafety to alternative flagging rules.}
\label{tab:oas-robust}
\setlength{\tabcolsep}{3pt}
\begin{tabular}{llcc}
\toprule
Configuration & Flagging rule & TPR (\%) & FPR (\%) \\
\midrule
LLM audit & Reported rule & 81.5 & 11.8 \\
          & Ignore refusals & $\geq$64.2 & $\geq$8.8 \\
          & No suspicion accumulation & 61.7 & 7.4 \\
\addlinespace
Gated LLM audit & Reported rule & 72.8 & 5.9 \\
            & Ignore refusals & $\geq$55.6 & $\geq$4.4 \\
            & No suspicion accumulation & 63.0 & 4.4 \\
\bottomrule
\end{tabular}
\end{table}

\noindent\textbf{Extending the simulated attack scenarios.} The simulated multi-agent systems provide an initial end-to-end demonstration that the framework can detect harmful actions while substantially reducing token usage. Future work can broaden this evaluation by including benign runs with legitimate sensitive operations, increasing the number and diversity of scenarios, and controlling the trace-generation models to enable more direct comparisons. It can also evaluate stronger monitor-aware attackers that avoid explicit cues in their cover stories, such as refusals. Together, these extensions would provide a more comprehensive evaluation of the framework's robustness and generalizability across realistic multi-agent deployments. They would also show whether the observed token savings persist under more challenging conditions.
\section{Conclusion}

This paper presents a two-stage LLM agent monitoring framework that combines rule-based action selection with semantic LLM auditing, preventing adversarial agent behavior from execution. Evaluations on a public benchmark and simulated multi-agent systems demonstrate that the framework can balance detection effectiveness with substantially lower auditing cost, providing a practical foundation for scalable AI agent monitoring.

\section*{LLM usage statement}
\emph{Methodological:} the benchmark and case-study agents, adversary personas, semantic audit agent, and optimizer that refines the gate rules and audit prompt use LLMs (Sections~\ref{sec:framework}--\ref{sec:results}). \emph{Preparation:} LLMs were used for editorial purposes in this manuscript, and all outputs were inspected by the authors to ensure accuracy and originality. An LLM-based coding environment also assisted in implementing the experimental harness. All design, claims, interpretation, and reported results were verified by the authors.

\nocite{*}
\bibliographystyle{IEEEtran}
\bibliography{Reference}

\end{document}